# NOVEL TYPES OF ANTI-ECLOUD SURFACES

I. Montero, L. Aguilera, M.E. Dávila, *Instituto de Ciencia de Materiales de Madrid-CSIC. Spain*
V. Nistor, L. Galán, L.A. González, *Departamento de Física Aplicada. Universidad Autónoma de Madrid. Spain*, D. Raboso, *European Space Agency/ESTEC, Noordwijk, The Netherlands*,
P. Costa Pinto, M. Taborelli and F. Caspers, *CERN, Geneva, Switzerland*, U. Ulrich and D. Wolk *Tesat-Spacecom, Germany*

*Abstract*

In high power RF devices for space, secondary electron emission appears as the main parameter governing the multipactor effect and as well as the e-cloud in large accelerators. Critical experimental activities included development of coatings with low secondary electron emission yield (SEY) for steel (large accelerators) and aluminium (space applications). Coatings with surface roughness of high aspect ratio producing the so-call secondary emission suppression effect appear as the selected strategy. In this work a detailed study of the SEY of these technological coatings and also the experimental deposition methods (PVD and electrochemical) are presented. The coating-design approach selected for new low SEY coatings include rough metals (Ag, Au, Al), rough alloys (NEG), particulated and magnetized surfaces, and also graphene like coatings. It was found that surface roughness also mitigate the SEY deterioration due to aging processes.

## INTRODUCTION

Intense research effort is presently directed toward solving the technical challenges of mitigating the electron cloud and its adverse consequences in large particle accelerators [1]. In addition, the multipactor discharge is a severe problem which limits the maximum power of RF instrumentation in space missions, such as Earth observation and telecommunication satellites, especially in multi-carrier operation [2]. It is also a problem in microwave generators, and thermonuclear toroidal plasma devices. Due to its technological and economic importance, the European Space Agency (ESA), CERN, and other prestigious research centres in the world, have dedicated a large effort for decades to its resolution. The main purpose of our research is to develop innovative anti-multipactor coatings with secondary electron emission suppressed by surface roughness effects, making good use of micro- and nano-tecnology. That research has obtained preliminary results which indicate that these coatings could be the solution to the multipactor problem. There are different techniques to suppress the electron cloud in large accelerators including clearing electrodes, direct reduction of the secondary electron yield (SEY) by coatings or cleaning/conditioning the surface or by increasing the surface roughness. We use different preparation techniques to optimize many interdependent material properties, like surface morphology and we combine fundamental studies, experimental research, and computer simulations to investigate the different growth processes and properties.

The effect of the surface roughness on the secondary emission in the field free case has been studied [3]. A technique for mitigating multipactor by means of magnetic surface roughness was proposed [4]. In addition, the influence of a static, but spatially alternating magnetic field in the LHC cavity RF power coupler to avoid multipactor effects, was studied [5].

## EXPERIMENTAL PART

The coatings were prepared by deposition and surface treatment techniques such as evaporation, sputtering, ion implantation, laser irradiation, and electrochemical ones. The coatings of metals such as Au and Ag and metalloids such as TiN with specially structured surface were deposited directly on the samples or in a multi-layer structure or with template techniques including structured ceramic oxide layers. Analysis techniques used are SEY and EDC for the secondary emission; XPS, EDX, GDOES, RBS, and synchrotron-radiation UPS, for chemical, structural, and electronic structure analysis; and SEM and AFM microscopies for surface morphology analysis (roughness).

SEY of the studied coatings has been characterized for incidence angles in the range 0 – 50 deg and for impacting energies in the range of 0 – 5000 eV, by the parameters: SEY maximum and its corresponding primary energy, and the first crossover energies for SEY equal to 1. The sample can be rotated in front of the XPS electron spectrometer for measuring surface composition and cleanliness, and in front of the programmable electron gun for the SEY measurements. SEY tests were carried by recording the sample current to ground after biasing the sample (-30 eV). The measurements were made via computer-controlled data acquisition; the sample was connected to a precision electrometer. The primary beam current was measured by a Faraday cup and several reference samples attached to the testing facility. The SEY is defined as $\sigma = (I_0 - I_s)/I_0$, where $I_0$ is the primary current and $I_s$ the sample one. The current $I_0$ is always negative, while $I_s$ can be positive or negative depending on the SEY values of the sample.

## ANTI-MULTIPACTOR COATINGS

Anti-multipactor coatings should have good surface electrical conductivity for avoiding RF losses, large resistance to air exposure, and low secondary emission. The main anti-multipactor coatings are the following:
i) *Alodine* is the typical chromate conversion coating for

aluminium used in space applications. However, it surface electrical conductivity is not sufficiently good.

ii) *Light transition metal nitrides, carbides, and suboxides*. These coatings are based on compounds with metallic character for low SEY while having as well relatively strong ionic-covalent bonding for chemical stability. TiN combines the strongest bond with the best electrical conductivity. In spite of this, all of them slowly degrade in air due to oxidation, water adsorption and other contaminations not well understood which make necessary surface conditioning. These have thoroughly been studied in the particle accelerators community. We have also studied several coatings of these kinds: TiN, VN, CrN, NbN, TiC, CN, CrSi, a-C … [6,7].

iii) *Light transition metal getter alloys (NEG)*. These coatings base their excellent performance on their surface self-cleaning mechanism by diffusion and solution of surface contaminants into the bulk by gentle heat treatments (activation).

iv) *Deeply surface-structured metals*. In this coating class the "SEY suppression effect" is a consequence of the strong surface roughness. This effect was known since many decades and recently, has fully been explained [8]. It affects more to primary electrons of lower energies, just the most important part of SEY for the multipactor discharge. This effect is more pronounced for materials with higher first cross-over energy. It is a classical macroscopic effect (no quantum physics) and it has been observed for sizes from 50 nm to several millimetres [8]. Our results also provide direct evidence of the reduction of SEY using micro-structured magnetized surfaces. The local magnetic field of these surfaces can modify the secondary electron trajectories making them to return back to the surface so that they are more easily reabsorbed.

## RESULTS AND DISCUSSION

The titanium nitride (TiN) coatings have good properties to prevent multipactor effect in RF wave guides as well as they can solve the multipacting problems of high power MW in particle accelerators. There is active research in this field. The difficulty in obtaining a TiN coating to prevent multipactor lies in the poor stability of the SEY coefficient after long exposures to air, or to contamination in the vacuum chamber during subsystem and satellite qualification tests and further "conditioning", see Fig.1.

Deterioration of TiN SEY properties because of long exposure to air can be recovered by different treatments (ion/electron bombardment or surface conditioning by vacuum heat treatments) once the surface is under vacuum for operation. However, TiN surface conditioning treatments are impossible or impractical in space applications. Figures 2 (a) and (b) show the SEM (Scanning Electron Microscope) images of a bilayer coating: (a) gold coated anomag (corrosion protection coating for Mg) on magnesium and (b) multipactor test results along with MEST (Multipactor Electron Simulation Tool) simulations, respectively. MEST works with a simplified model of a waveguide, where a radiofrequency potential is applied to the parallel plates. The space between the plates is assumed to be a vacuum, except for the presence of a number of free electrons. The simulator detects the multipactor discharge and generates V-f·d maps. [9]. We can observe that the highest breakdown voltage was obtained for gold-coated anomag. MEST result of gold after air exposure has been included for comparison purposes.

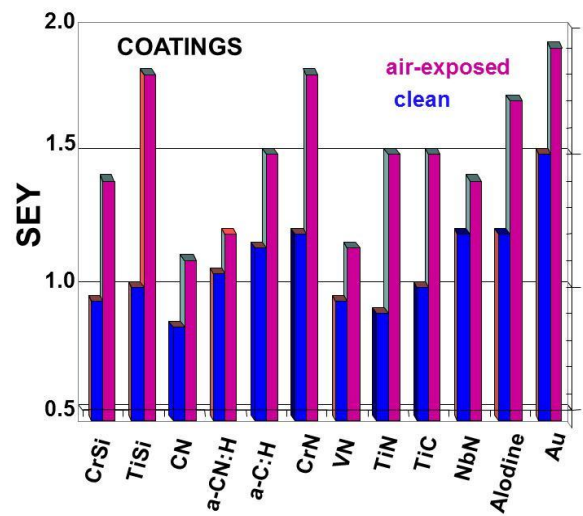

Figure 1.- Influence of air exposition on SEY of different selected coatings.

Figures 2 (a) and (b) show the SEM (Scanning Electron Microscope) images of a bilayer coating: (a) gold coated anomag (corrosion protection coating for Mg) on magnesium and (b) multipactor test results along with MEST (Multipactor Electron Simulation Tool) simulations, respectively. MEST works with a simplified model of a waveguide, where a radiofrequency potential is applied to the parallel plates. The space between the plates is assumed to be a vacuum, except for the presence of a number of free electrons. The simulator detects the multipactor discharge and generates V-f·d maps. [9]. We can observe that the highest breakdown voltage was obtained for gold-coated anomag. MEST result of gold after air exposure has been included for comparison purposes.

There are different experimental methods to obtain surface roughness of high aspect ratio: chemical etching process, ion bombardment under special conditions (pressure, incidence angle...). Also, copper oxide nanowires grown on copper foil present very low SEY maximum value (0.6).

Figures 3 and 4 show the SEM (Scanning Electron Microscope) images of Ag plating, Ar ion etched assisted by Mo sputtering deposition, and a multilayer coating, gold-coated rough silver, respectively. During low-energy Ar ion bombardment of the flat silver plate at room temperature, the different sputtering rate of molybdenum compared to silver induces a suitable surface roughness. The resulting molybdenum concentration on the surface is

<1%. Surface roughness of controlled aspect ratio can be obtained.

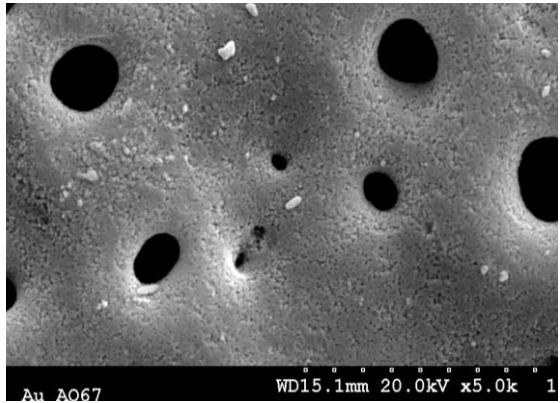

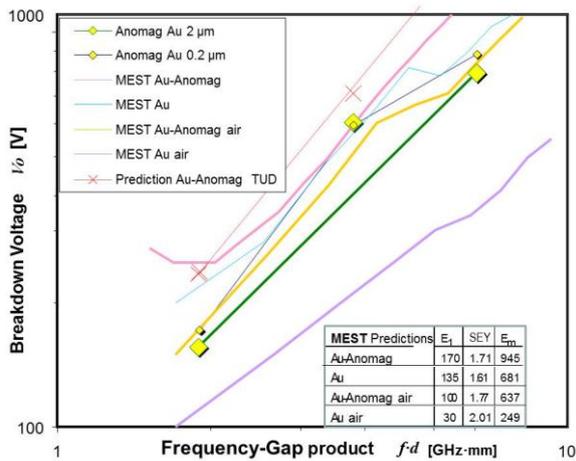

Figure 2. (a) SEM image of gold coated Anomag and (b) Multipactor test results and MEST simulations of gold coated Anomag.

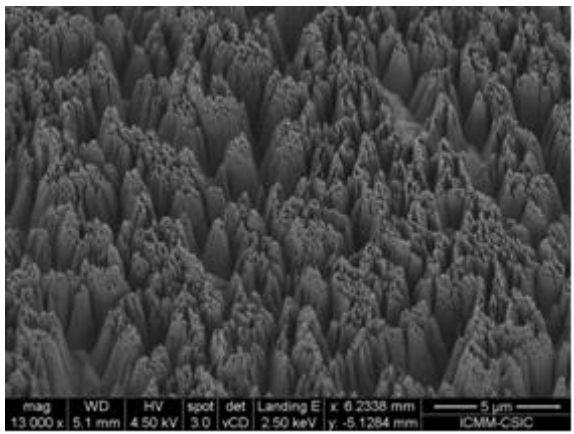

Figures 3. SEM (Scanning Electron Microscope) images of Ag plating treated by Ar ion etching assisted by Mo sputtering.

SEY results of rough silver before and after gold deposition are shown in Fig.4; the SEY curves of flat silver and flat gold have been included for comparison purposes. A strong decrease of SEY is mainly observed at lower primary energies.

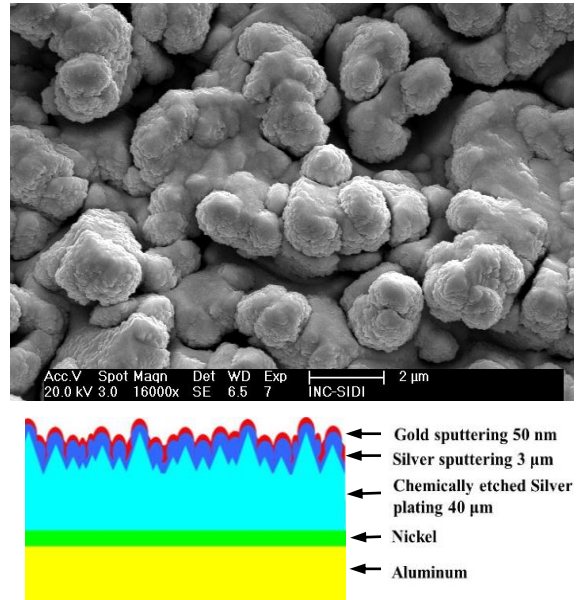

Figure 4. SEM image of the multilayer coating of gold-coated chemically-etched silver (a) and scheme of this multilayer coating (b).

In Fig.4 we can also observed the high first-cross over energy value, $E_1$ = 400 eV, and also the low value of the SEY maximum, 1.15, of gold coated rough silver.

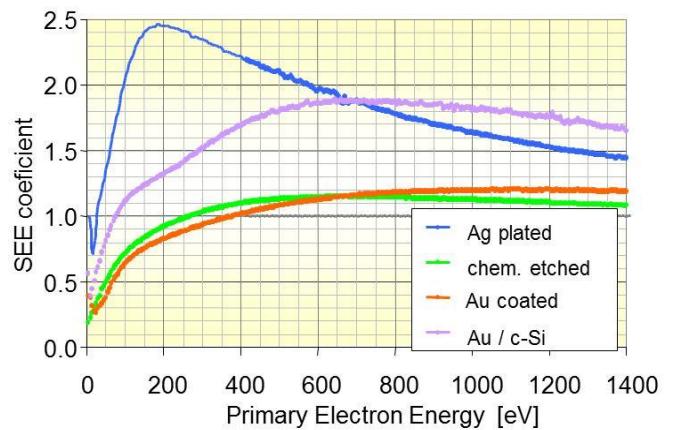

Figure 5. SEY curves of gold coated rough silver before and after chemical etching process. Gold coated flat c-Si has been included for comparison purposes.

The mean surface roughness values (Ra) determined with profilometry of etched aluminium at different etching times (60, 120 and 150 s), and the corresponding SEM images of the transversal sections of the etched Al are shown in Fig.6.

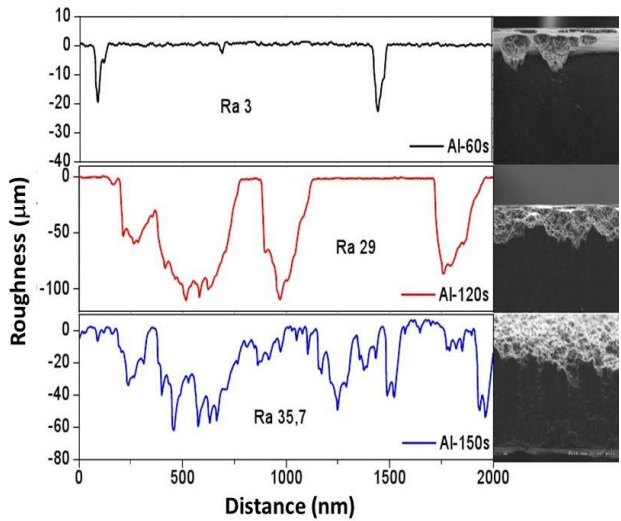

Figure 6. Profilometry results of the etched aluminium, Mean surface roughness values (Ra) and the corresponding SEM images of their transversal sections.

Apart from XPS and EDX, NEG coated rough aluminium was also analyzed by GDOES. Figure 7 shows the glow discharge optical emission spectroscopy depth profile of NEG/r-Al. In this depth profile analysis the depth resolution mainly depends on the roughening induced during the sputtering and the resulting crater geometry. However, the existence of a surface roughness produces a strong broadening of the interface width.

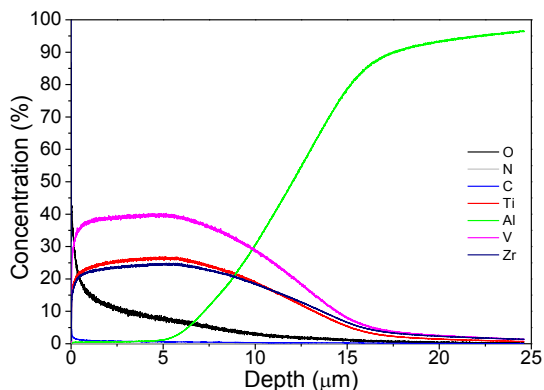

Figure 7. GDOES depth profile results of the NEG coated etched aluminium, NEG/r-Al.

Figure 8 shows the secondary emission yield of NEG/r-Al after activation process as a function of the primary energy and aluminum etching time. A noticeable decrease of SEY is observed even at high primary energy. The positive effect of the additional etching is also clear here.

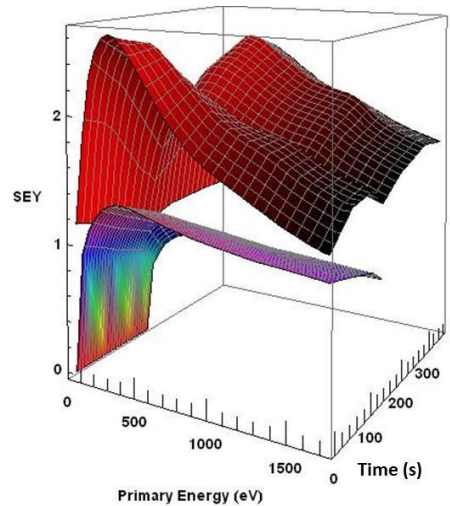

Figure 8. SEY curves of NEG/r-Al as a function of primary energy and aluminum etching time. Lower surface, after additional etching.

## SUMMARY AND CONCLUSIONS

To prevent multipactor effect, coatings with low secondary emission, stability after exposure to air, and low surface electrical conductivity can only be attainable by means of a suppression or absorption effect due to multiple incidences in rough surfaces with special surface morphology with large aspect ratio.

## ACKNOWLEDGMENTS

This work was supported by Projects No. AYA2009-14736-C02-01 and 02 (space programme) of MICIIN of Spain, ITI programme of ESA, and CERN support and cooperation.

## REFERENCES

[1] http://ab-abp-rlc.web.cern.ch/ab-abp-rlc-ecloud/
[2] http://multipactor.esa.int/whatis.html
[3] L. Galán, V. Nistor, I. Montero, L. Aguilera, D. Wolk, U. Wochner and D. Raboso. Proceedings of MULCOPIM' 2008. ESTEC-ESA, Noordwijk, The Netherlands (2008).
[4] F. Caspers, E.Montesinos, I.Montero, V. E. Boria, C. Ernst, D.Raboso, L. Galan, B.Gimeno, W.Bruns, C. Vicente. Proc. 23rd Particle Accelerator Conference, PAC (2009).
[5] I.Montero, F.Caspers, L.Aguilera, L.Galán, D.Raboso and E. Montesinos. Proceedings of IPAC'10, Kyoto, Japan, Kyoto, Japan, (2010), pp.TUPEA077.
[6] C. Yin Vallgren, S. Calatroni, P.Costa Pinto, A.Kuzucan, H.Neupert, M.Taborelli, Proceedings of IPAC2011, San Sebastián, Spain.
[7] F. Caspers, E.Montesinos, I.Montero, V. E. Boria, C. Ernst, D.Raboso, L. Galan, B.Gimeno, W.Bruns, C. Vicente. Proc. 23rd Particle Accelerator Conference, PAC (2009).
[8] M Pivi, F K King, R E Kirby, T O Raubenheimer, G Stupakov, F Le Pimpec. J. Appl. Phys., 104, 104904-10 (2008).
[9] J.de Lara, F. Perez, M. Alfonseca, L. Galan, I. Montero, E. Roman, D.R. Garcia-Baquero IEEE Trans. Plasma Sci., vol. 34, no. 2 (2006).